\documentclass[oldversion]{./aa}
\usepackage{graphicx}
\usepackage{txfonts}
\usepackage{natbib}
\renewcommand{\tabcolsep}{1.4mm}
\bibpunct{(}{)}{;}{a}{}{,} 

\def\ecs{erg~cm$^{-2}$s$^{-1}$}
\def\lum{erg~s$^{-1}$}
\def\bron{GS~1826-24}

\begin{document}

\title{Long tails on thermonuclear X-ray bursts from neutron stars: a
  signature of inward heating?}

\titlerunning{Long tails on thermonuclear X-ray bursts}
\authorrunning{J.J.M. in 't Zand, L. Keek, A. Cumming et al.}

\author{J.J.M.~in~'t~Zand\inst{1}, L. Keek\inst{1,2}, A. Cumming\inst{3},
  A. Heger\inst{4}, J. Homan\inst{5} \& M.~M\'{e}ndez\inst{6}}

\offprints{J.J.M. in 't Zand, email {\tt jeanz@sron.nl}}

\institute{     SRON Netherlands Institute for Space Research, Sorbonnelaan 2,
                3584 CA Utrecht, the Netherlands
	 \and
                Astronomical Institute, Utrecht University, P.O. Box 80000,
                3508 TA Utrecht, the Netherlands
         \and
                Physics Department, McGill University, 3600 Rue University, 
                Montreal, QC, H3A 2T8, Canada
         \and
                School of Physics \& Astronomy, University of Minnesota,
                Twin Cities, Minneapolis, MN 55455, U.S.A.
	 \and
                MIT Kavli Institute for Astrophysics and Space Research,
                70 Vassar Street, Cambridge, MA 02139, U.S.A.
         \and
                Kapteyn Astronomical Institute, Groningen University,
                9700 AV Groningen, the Netherlands
          }

\date{1st revised version (\today)}

\abstract{We report the discovery of one-hour long tails on the
  few-minutes long X-ray bursts from the `clocked burster' \bron. We
  propose that the tails are due to enduring thermal radiation from
  the neutron star envelope. The enduring emission can be explained by
  cooling of deeper NS layers which were heated up through inward
  conduction of heat produced in the thermonuclear shell flash
  responsible for the burst. Similar, though somewhat shorter, tails
  are seen in bursts from EXO~0748-676 and 4U 1728-34. Only a small
  amount of cooling is detected in all these tails. This is either due
  to compton up scattering of the tail photons or, more likely, to a
  NS that is already fairly hot due to other, stable, nuclear
  processes.

\keywords{X-rays: binaries -- X-rays: bursts -- accretion, accretion
  disks -- stars: neutron -- X-rays: individual: \bron, EXO~0748-676,
  4U 1728-34~=~GX354-0}}

\maketitle

\section{Introduction}
\label{intro}

Type-I X-ray bursts, or X-ray bursts in short, result from
thermonuclear shell flashes of hydrogen and helium on neutron stars
(NSs). The fuel is accreted from a Roche-lobe-filling companion
star. As the accretion of this material progresses, the pressure at
the bottom of the accreted layer rises to ignition conditions for
thermonuclear fusion processes like the (hot, $\beta$-decay limited)
CNO cycle, triple-$\alpha$ process, $\alpha$-proton capture and the
($\beta$-decay limited) rapid-proton capture (rp) process. In general,
the fuel is burnt within 1 s in the top few meters of the NS and
temperatures momentarily reach values as high as a few GK. The layer
is covered by a non-burning layer, on top of which is the
photosphere. What one measures is the cooling flux passing through the
photosphere, with temperatures peaking at about 30 MK. The burst
duration is primarily determined by the time it takes to cool the
shell. For H-rich flashes, the duration is further lengthened due to
prolonged nuclear burning through the rp process. The burst duration
may range from a few seconds to a few hundred seconds for the large
majority of X-ray bursts. Exceptional durations come from very thick
helium layers on relatively cold NSs in hydrogen-poor ultracompact
X-ray binaries \citep[i.e., up to thousands of seconds; e.g.,
][]{zan05a, cum06, zan07} and flashes of very thick carbon shells
\citep[`superbursts'; e.g.,][]{cor00,cum01,stro02}. For reviews on
X-ray bursts and further references, we refer to \cite{lew93},
\cite{bil98} and \cite{stroh06}.

Sometimes very long tails are seen in X-ray bursts that are not
related to the aforementioned long bursts. For instance, \cite{che06}
discuss a peculiar burst from GX 3+1. Since the advent of X-ray
astronomy, about 100 bursts have been detected from this source
\citep[e.g.,][]{hartog03} and all are shorter than a few tens of
seconds except for this peculiar burst. It has a prolonged tail that
starts at about 25\% of the ordinary burst peak flux and decays with
an e-folding decay time of 1110~s (for photons between 3 and 6
keV). \cite{che06} hypothesize that the tail is due to rp capture of a
rich hydrogen mixture that became available after the accretion rate
dropped below the threshold where hydrogen is burnt in a stable manner
(GX 3+1 was in a 10-yr minimum at about the time of the
burst). However, the time scales of the slowest $\beta$ decays
expected in X-ray bursts are at least one order of magnitude shorter
than 1110~s \citep[e.g.,][]{fis08}. A few similar cases (i.e., from
systems that are clearly not ultracompact X-ray binaries) are
described in the literature, most notably in \cite{cze87} and
\cite{got97}. There is a clear duality in the time profile of these
bursts: they start with an ordinary short-lived burst, followed by a
$10^{2-3}$~s long tail without an unambiguous cooling signature. The
long tail starts off quite brightly in these bursts, at a few tens of
percents of the burst peak flux. Since the nature of these long tails
is not well established, it is worthwhile exploring whether there are
long tails that start off at a lower fraction of the burst peak (i.e.,
percents instead of tens of percents).

Burst studies usually concentrate on the brightest parts, roughly
above 1\% of the peak flux \citep[which is often close to the
  Eddington limit, see][]{gal06}, and for a good reason. Most bursts
come from prolific bursters with mass accretion rates above the same
1\% level. Since accretion is notoriously variable, this makes
disentangling burst radiation from accretion radiation difficult at
fluxes below 1\% of the Eddington limit.

Exceptions are bursts from NSs that accrete at rates below 1\% of the
Eddington limit. This pertains to most persistently accreting
ultracompact X-ray binaries \citep[UCXBs; a nice example of a burst
  that could be studied with Swift to very deep levels originates in
  A~1246-588;][]{zan08} and bursts from transients whose accretion
rate has dwindled down to low but non-zero values \citep[e.g.,
][]{zan03}. These exceptions show tails that are natural extensions of
the decays of the bright parts of the same X-ray bursts. In other
words: there is no prompt/tail duality.

Despite the fact that it is difficult to study bursts at sub-1\%
levels in fast-accreting bursters, the situation is sometimes not
desperate. This paper presents a study of the unique burster
\bron. The accretion is very stable in this source. The variability on
a time scale of the burst recurrence time (few hours) is about 2\%
rms, so that the recurrence time from burst to burst is relatively
stable as are the burst peak flux and profile
\citep{ube99,coc01,gal04,heg07}. This behavior earned it nicknames
such as `the clocked burster' \citep{ube99} and `the textbook burster'
\citep{bil00}. \bron\ is also notable for a high-energy component of
the burst emission \citep[i.e., above 30 keV where negigible amounts
  of black body emission are expected;][]{zan99a}. Because of its
stable bursting behavior and accretion rate, \bron\ is excellently
suited for low-level burst flux studies. The most recent distance
determination is $6.07\pm0.18$~kpc \citep[for isotropic burst
  radiation;][]{heg07}.

The existence of long tails in \bron\ was already implied in
\cite{zan99a} and \cite{tho05}. The latter work concentrated on the
persistent spectrum as measured with Chandra and RXTE. This emission
could be successfully modeled by a combination of two comptonization
components \citep[see also][]{tho08} due to the presence of hot
plasmas in the immediate neighborhood of the NS. Bursts were studied
as well, and their spectra could be modeled by a black body component
plus the variation of one of the persistent comptonization
components. This is in line with the high-energy burst emission seen
by \cite{zan99a} and in contrast with the so-called `standard'
modeling of burst spectra where the burst emission is modeled solely
by one black body component and the persistent spectrum is unaffected
by the burst emission. A simple physical interpretation of the change
of one comptonization component would be that a hot plasma up-scatters
some of the burst thermal photons to higher energies and itself is
cooled down by the soft photons. This model was applied to 1000~s of
burst data in \cite{tho05}. The final 850 s of the data were modeled
by comptonization only.

In this paper we make a study of the long tail in \bron\ employing all
RXTE data available and cross checking with XMM-Newton data. Thus, we
obtain a superior statistical quality and are able to probe the tail
longer than \cite{tho05}. We supplement this analysis with briefer
investigations of two other prolific bursters, EXO 0748-676 and 4U
1728-34. We propose an explanation for the tail.

\section{Observations}
\label{obs}

\begin{table}
\caption[]{Selection of X-ray bursts from \bron\ studied here. For more
details, see \cite{gal06}\label{tab1}}
\begin{tabular}{llllc}
\hline\hline
MJD & ObsID & Times & Active & $\Delta^a$ \\
    &       & covered wrt & PCUs  & (\arcmin) \\
    &       & burst start&   &  \\
    &       & time (sec)$^b$    & &  \\
\hline
50971.23019          & 30054-04-02-01  &  -2600/+1200 & 0,1,2,3,4  & 0.0 \\
50971.70032          & 30054-04-02-000 &  -2700/+900  & 0,1,2,3,4  & 0.0 \\
50988.82503          & 30060-03-01-01  &  -1700/+1800 & 0,1,2,3    & 0.0 \\
51725.87648          & 50035-01-02-00  &  -3000/+200  & 0,2,3,4    & 0.5 \\ 
51726.71988$^{c,p,t}$ & 50035-01-02-02  &  -1200/+2100 & 0,1,2,3    & 0.5 \\
51728.77338$^{c,p,t}$ & 50035-01-02-04  &  -4000/+2600 & 0,1,2,3    & 0.5 \\
51814.34546$^{c,p,t}$ & 50035-01-03-10  &  -2200/+1500 & 0,2        & 0.5 \\
52484.41775$^{c,p,t}$ & 70044-01-01-000 &  -4000/+2200 & 0,2,3      & 0.0 \\
52484.56684$^c$      & 70044-01-01-00  &  -2400/+700  & 0,2,3      & 0.0 \\
52485.00672$^{c,p,t}$ & 70044-01-01-02  &  -4000/+2600 & 0,2,3      & 0.0 \\
52537.22347          & 70025-01-01-00  &  -4000/+2400 & 0,1,2,3    & 0.0 \\
52738.47686$^{c,p,t}$ & 80048-01-01-04  &  -4000/+3200 & 0,2,3      & 0.0 \\
52738.74636$^{c,p,t}$ & 80048-01-01-07  &  -4000/+2700 & 0,1,2,3    & 0.0 \\
52820.56256$^{c,p,t}$ & 80049-01-01-00  &  -1100/+2000 & 0,2,3,4    & 0.0 \\
52893.05439          & 80105-11-01-00  &  -4000/+3000 & 0,2,3      & 0.0 \\
53205.19659$^{c,p,t}$ & 70044-01-02-00  &  -1900/+1500 & 0,2,3      & 0.0 \\
53207.07579          & 90043-01-01-01  &  -4000/+3600 & 0,2        & 2.5 \\
53207.21911$^c$      & 90043-01-01-01  &  -4000/+1400 & 0,2,3      & 2.5 \\
53206.49150$^{c,p,t}$ & 90043-01-01-020 &  -4000/+2800 & 0,2,3      & 2.5 \\
53206.63615$^{c,p,t}$ & 90043-01-01-02  &  -4000/+1900 & 0,2,3      & 2.5 \\
53956.08413$^{p,t}$  & 92031-01-01-000 &  -4000/+2200 & 0,2,4      & 13.8 \\
53956.22695$^{p,t}$  & 92031-01-01-000 &  -2100/+1100 & 0,1,2      & 13.8 \\
53957.06785$^{p,t}$  & 92031-01-01-010 &  -4000/+2000 & 0,2,3      & 13.8 \\
53957.20458$^{p,t}$  & 92031-01-01-010 &  -1800/+1600 & 0,2,3      & 13.8 \\
53959.15478$^{p,t}$  & 92031-01-02-000 &  -4000/+2900 & 1,2        & 13.8 \\
54167.72354$^{c,p,t}$ & 91017-01-01-000 &  -1900/+1100 & 1,2        & 0.0 \\
54168.31443$^{c,p,t}$ & 91017-01-02-03  &  -2000/+1600 & 2          & 0.0 \\
54168.75518$^{c,p,t}$ & 91017-01-02-010 &  -4000/+2500 & 0,2,3      & 0.0 \\
54168.90229$^{c,p,t}$ & 91017-01-02-010 &  -1900/+1600 & 0,2        & 0.0 \\
\hline
\end{tabular}
$^a$ Off-axis angle;
$^b$These values are rounded off to an integer times 100 s and may
include gaps. $^c$The spectra for these ObsIDs were used to
generate average time-resolved spectra (cf.,
Fig.~\ref{figxtespectra}); $^p$Data were used to calculate average
pre-burst power density spectrum (Fig.~\ref{figpds}); $^t$Data were
used to calculate average tail power density spectrum.
\end{table}

We use data from the Proportional Counter Array (PCA) on the Rossi
X-ray Timing Explorer (RXTE) and the European Photon Imaging p-n
junction camera (EPIC-pn) on the XMM-Newton observatory. These data
sets were chosen because they are complementary in photon energy
bandpass (2-60 and 0.1-12 keV, respectively) with similar
sensitivities and because they detected more than 10 X-ray bursts
each.

The PCA comprises 5 identical proportional counter units (PCUs) with a
total net collecting detector area of about 8000~cm$^2$ in the 2-60
keV band \citep[3-20 keV is well calibrated;][]{jah06}. During any
observation any number of PCUs between 1 and 5 are on. The PCA is a
non-imaging device with a spectral resolution of 20\% (full width at
half maximum; FWHM) and a 2 degree wide field of view (full width to
zero response). There are no other bright X-ray sources in the field
around \bron.

RXTE is particularly suited for the study of low-mass X-ray binaries,
and since its 1995 launch has accumulated unprecedented exposure times
on many X-ray bursters, one of them being \bron. \cite{gal06} compiled
a catalog of all bursts detected between Feb 8, 1996, and June 3,
2007.  The total PCA exposure time on \bron\ over 127 observations in
this time frame is 929~ks. 65 bursts were detected between Nov 5,
1997, and March 10, 2007, and the mean burst rate was
0.25~hr$^{-1}$. None of the bursts exhibit evidence for photospheric
radius expansion, so that the peak luminosity must have been below the
Eddington limit. The bolometric absorption-corrected peak flux ranged
between $23.5\pm0.8$ and $(29.6\pm0.8)\times10^{-9}$~\ecs\ and the
fluence between $0.923\pm0.016$ and
$(1.059\pm0.004)\times10^{-6}$~erg~cm$^{-2}$ \citep{gal06}. The decay
can be modeled by two subsequent exponential decays. The e-folding
decay times ranged between 12.9 and 23.3 s for the first decay and
between 40.5 and 57.2 s for the second decay. The average time scale
for the decay (defined as the fluence divided by the peak flux) is
between 30 and 45 s, which implies that the 0.1\% level is reached
within 400 s.

XMM-Newton observed \bron\ on two occasions: starting on April 6,
2003, for 108 ks and April 8, 2003, for 92 ks. A complete account of
these observations is provided in \cite{kon07}. Nine X-ray bursts were
detected in the first and seven in the second observation. The final
bursts in each observation suffered from high background levels and
were excluded from the analysis. All X-ray detectors were on, but we
concentrate on the 0.1-12 keV EPIC-pn measurements because that
instrument \citep{stru01} is by far the most sensitive for our
analysis. It has an effective area that ranges between
1000-3000~cm$^2$ for 0.5-2 keV and 900~cm$^2$ at 2-6 keV. The spectral
resolution is 2.5\% (FWHM) at 6 keV. The instrument was switched to
Fast Timing mode, implying that the central CCD (of the 12 available),
encompassing 13\farcm6$\times$4\farcm4 of the field of view, is read
out 1-dimensionally (along the 4\farcm4 side) at a 1.5~ms
resolution. Source photons were extracted between {\tt RAWX} values of
30 and 45, with pixel patterns below 5 and grade 0; background photons
were extracted between {\tt RAWX=}10 and 25. We refer to \cite{kon07}
for further details, noting that SAS version 7.1.0 was employed for
our data analysis.

\begin{figure}[t]
\includegraphics[height=\columnwidth,angle=270]{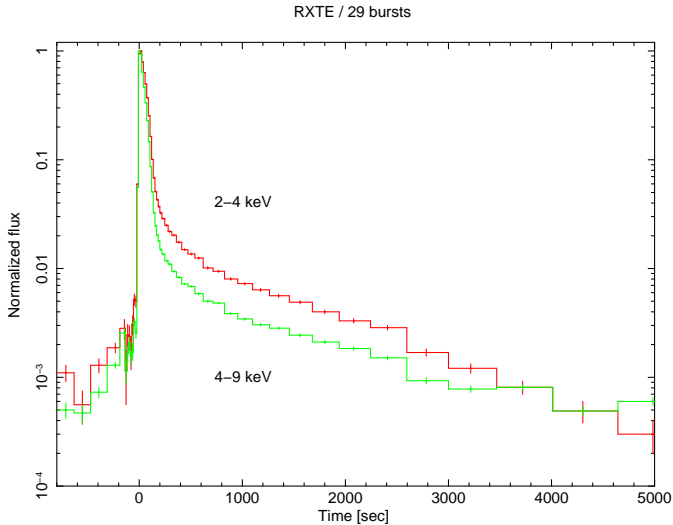}
\caption{Average time profile of 29 X-ray bursts from GS 1826-24 as
  measured with PCU2 on RXTE/PCA, in two photon-energy bandpasses and
  at logarithmically scaled resolution.\label{figxteprofiles}}
\end{figure}

\begin{figure}[t]
\includegraphics[height=\columnwidth,angle=270]{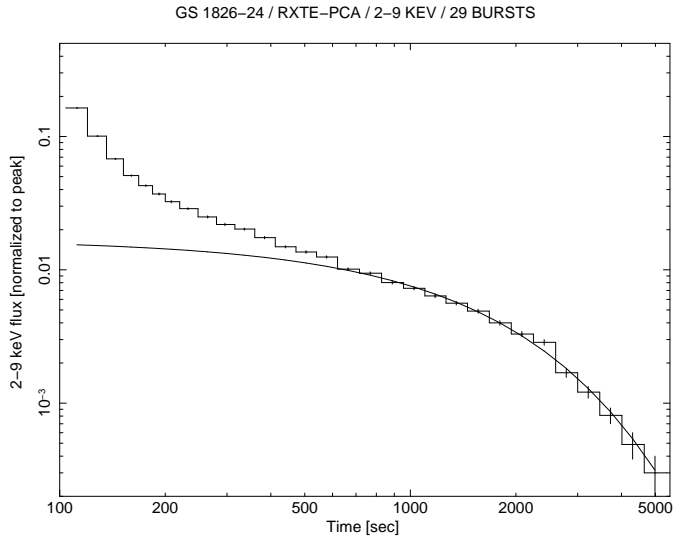}
\caption{Average time profile of 29 X-ray bursts from GS 1826-24 as
  measured with PCU2 on RXTE/PCA in 2-9 keV and at logarithmically
  scaled resolution, together with an exponential decay function
  (smooth curve) as fitted between 600 and 5500 s ($\tau=1252$~s; see
  Table~\ref{tab2})\label{figxtefit}}
\end{figure}

\begin{figure}[t]
\includegraphics[height=\columnwidth,angle=270]{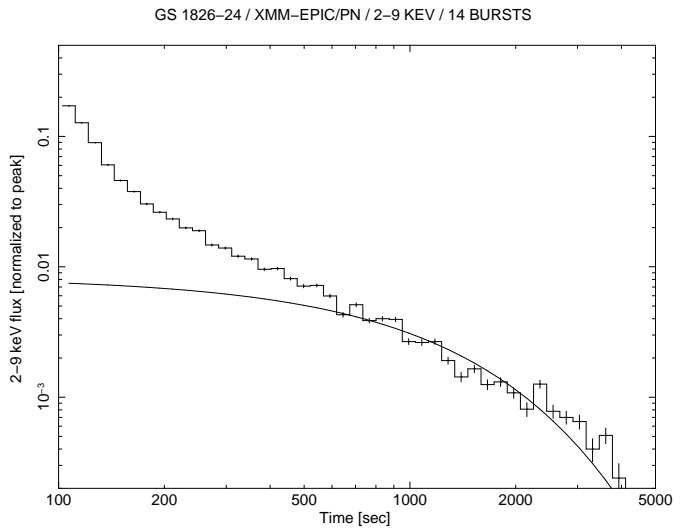}
\caption{Average time profile of 14 X-ray bursts from GS 1826-24 as
  measured with XMM/EPIC, in one bandpass and at a logarithmically
  scaled resolution. The count rate scaling is different from
  Fig.~\ref{figxteprofiles} due to the different time resolution at
  the peak and the occurrence of saturation effects in the XMM-Newton
  data. The smooth curve shows an exponential function, fitted between
  600 and 3000 s. The decay time was found to be
  $\tau=1006\pm26$~s. The goodness-of-fit is formally unacceptable
  ($\chi^2/\nu=179/22$), so this result is only
  indicative. \label{figxmmprofile}}
\end{figure}

\section{Data analysis}
\label{ana}

We selected 29 out of the 65 PCA-detected bursts that have data
available between 4000 and 1000 s before the burst start and between
400 and 3000 s afterwards, see Table~\ref{tab1}. The pre-burst
interval was based on the general trend that the flux was lowest there
and should provide the best estimate of the accretion flux. We
processed the data of the 29 bursts to an average light curve as
follows.  Taking the `standard-products' light curves as starting
point (these are data from PCU2 in 5 energy bands, corrected for
particle-induced background and with a resolution of 16 s) we
determined the pre-burst flux level from data between 4000 and 1000 s
prior to the burst (note that data gaps are common in this 3000~s time
frame), subtracted that from all flux measurements, co-aligned the
data at the start time of the burst \citep[defined as the time when
  the photon flux exceeds 25\% of the peak flux;][]{gal06} and
averaged all bursts. Figure~\ref{figxteprofiles} shows the resulting
profiles in 2 to 4 and 4 to 9 keV, normalized to the peak flux. Apart
from the well-known burst profile with a duration of about 400 s, it
shows the striking appearance of an additional burst component that
lasts approximately ten times longer at flux levels between 10$^{-3}$
and 10$^{-2}$ times the burst peak value. A comparison with higher
time-resolution data shows that the averaging of the peak in 16~s time
bins lowers the peak flux by approximately a factor of 2. While the
initial decay of the profile shows the classical cooling of the bursts
(the high-energy flux decaying faster than the low-energy flux), the
slow decay does not show obvious cooling.

\begin{figure}[t]
\includegraphics[width=\columnwidth]{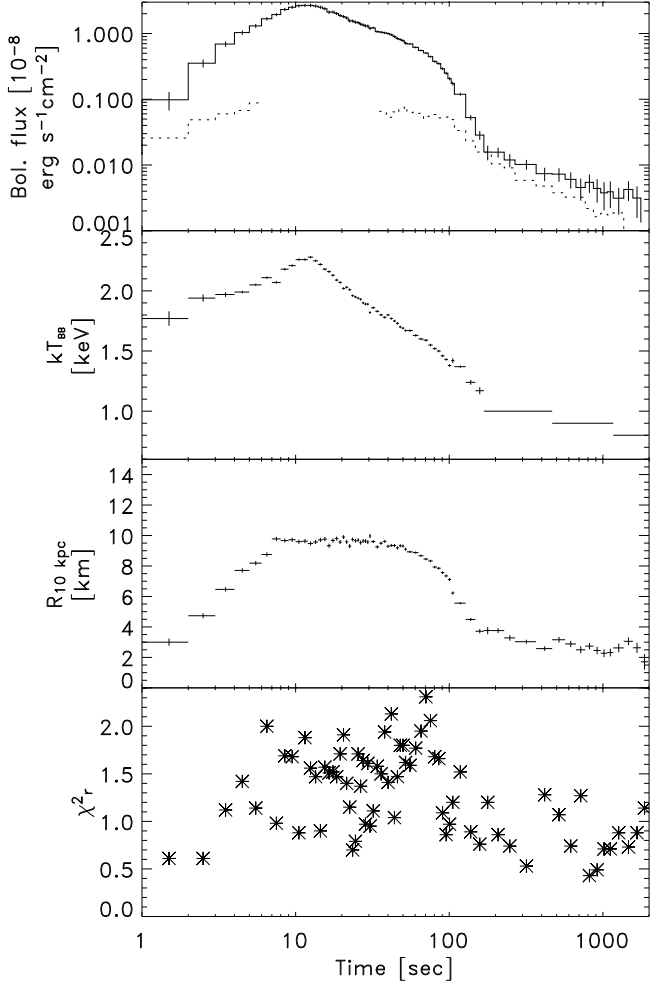}
\caption{Modeling results for time-resolved PCA burst spectra of
  average of 17 bursts. The model consists of a black body and
  power-law component absorbed by a fixed column of $N_{\rm H}=3.1
  \times 10^{21}$~cm$^{-2}$. The panels show from top to bottom: the
  bolometric flux for the black body component (solid curve) and the
  unabsorbed 3-20 keV flux found for the power-law component (dashed
  curve; gaps indicate times when the power law is not detected), the
  black body temperature, the emission area in terms of the radius of
  a sphere at a canonical distance of 10 kpc, and the goodness-of-fit
  in terms of reduced $\chi^2$ (the number of degrees of freedom is 17
  or 18). The black body temperature is, beyond 150 s, fixed to the
  values found in the XMM-Newton data.\label{figxtespectra}}
\end{figure}

\begin{figure}[t]
\includegraphics[width=\columnwidth]{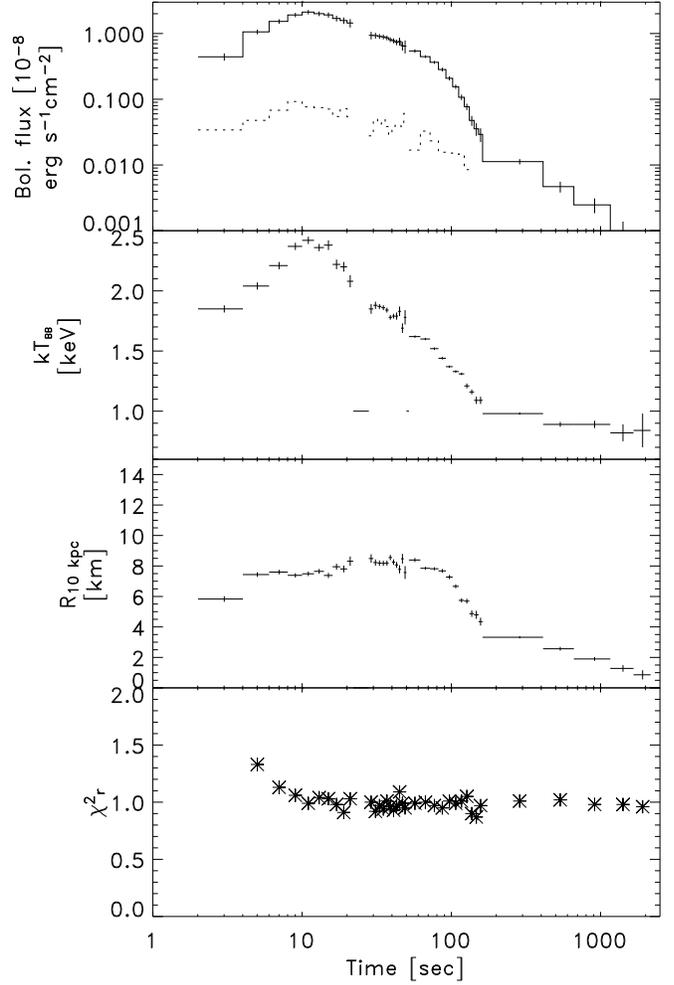}
\caption{Modeling results for time-resolved 0.7-10 keV burst spectra
  of 14 bursts detected XMM-Newton observation. Spectral channels were
  binned so that each bin contains at least 15 photons. The number of
  degrees of freedom ranges between roughly 500 and 2000 (for 2
  spectra, one for each observation) which explains the small scatter
  in $\chi^2_\nu$. The anomalous values at 20~s are due to data drop
  outs due to full data buffers from high photon rates
  \citep[see][]{kon07}.\label{figxmmspectralfits}}
\end{figure}

The data can be described satisfactorily by an exponential decay
function between 600 and 5500 s. For the 2-9 keV time profile the fit
is shown in Fig.~\ref{figxtefit}. the decay time is $1252\pm25$~s
($\chi^2/\nu=16.9/13$). Resolved in the two bands the e-folding decay
times are 1261$\pm$29~s in 2-4 keV ($\chi^2/\nu=16.3/10$) and
1381$\pm$29 in 4-9 keV (as measured between 600 and 3500 s after burst
onset for lack of statistics beyond 3500 s; $\chi^2/\nu=53.4/10$).
Since it is expected that cooling by thermal/photon diffusion follows
a power law \citep{eic89} we fitted such a law and find, between 300
and 2300 s, a power law index of $-0.93\pm0.02$.

We sought verification of the long tail in data from the XMM-Newton
observations. Figure~\ref{figxmmprofile} shows the 2-9 keV light
curve, averaged over 14 out of the 16 X-ray bursts (leaving out the
last burst of each observation, being compromised by increased
background levels). The same long tail is seen as with RXTE. The
e-folding decay time is $\tau=1006\pm26$~s between 600 and 2400 s
which is similar as for the RXTE curve. A tail is also seen in data
below 2 keV, but the statistics are not so good to reveal it beyond
10$^3$ s. The e-folding decay time in this low-energy bandpass is
$524\pm172$~s between 400 and 1000~s, which at least shows that it is
not longer than for the 2-9 keV band.

\begin{figure}[t]
\includegraphics[height=\columnwidth,angle=270]{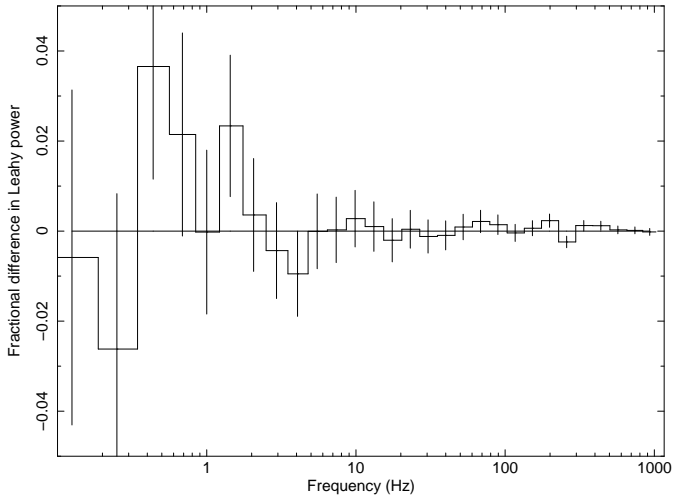}
\caption{Geometrically binned Fourier power density difference
  spectrum between pre-burst and tail data (in formula: tail pds minus
  pre-burst pds divided by pre-burst pds). \label{figpds}}
\end{figure}

\begin{figure}[t]
\includegraphics[width=\columnwidth,bb=64 36 378 515,clip]{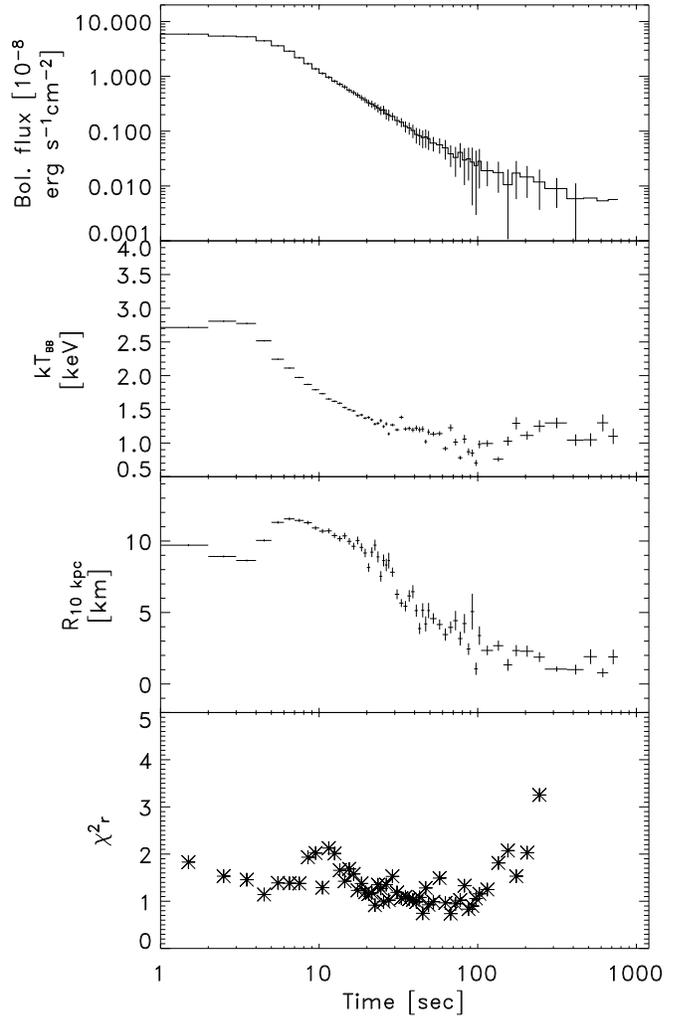}
\caption{Modeling results for time-resolved PCA burst spectra of
  average of 36 bursts from 4U 1728-34. The model consists of a black
  body and power-law component absorbed by a fixed column of $N_{\rm
    H}=2.6 \times 10^{22}$~cm$^{-2}$. The panels show from top to
  bottom: the bolometric flux for the black body component, the black
  body temperature, the emission area in terms of the radius of a
  sphere at a canonical distance of 10 kpc, and the
  goodness-of-fit. The data points beyond 250 s have large $\chi^2$
  values (outside range of bottom plot) and do not fit the model. This
  is most likely due to variability in accretion
  flux.\label{fig1728spec}}
\end{figure}

We performed time-resolved spectroscopy.  Spectra were accumulated for
17 bursts that have small off-axis angles in the PCA and have
identical detector-voltage settings (i.e., they are said to be in the
same `gain epoch'; see specification in Table~\ref{tab1}).  Only PCU2
data were employed to obtain a homogeneous data set. The time
resolution of the spectral extraction was chosen to vary between 1 s
early on in the burst and 200 s at the end. Each burst was divided in
72 time bins that are identical with respect to the burst start
times. Spectra were extracted from event mode data with s/w tool {\tt
  seextrct} for these time intervals and corrected for the
particle-induced background as determined with {\tt pcabackest} in the
same time interval. For each burst, a pre-burst spectrum was generated
from data available between 4000 and 1000 s prior to the burst, which
also was corrected for particle-induced background. Subsequently the
spectra for all 17 bursts were averaged in their respective time
frames, and the average pre-burst spectrum was subtracted from the 72
burst spectra. These spectra were, between 4 and 20 keV, modeled with
a simple absorbed black body function, employing a fixed absorption
column of $N_{\rm H}=3.1\times10^{21}$~cm$^{-2}$ (see XMM-Newton
analysis below). The photo-absorption cross sections were taken from
\cite{bal92} and the composition of the absorbing material from
\cite{wil00}.  About half of the spectra (in the brightest phase) turn
out not to be consistent with this model and an additional power-law
component with a fixed index (equal to that of the pre-burst data) was
included in the model. The fitted values for the various black body
parameters are shown in Fig.~\ref{figxtespectra}. All time intervals
are well fitted. During the first 100~s the black body parameters are
as expected for an X-ray burst: a temperature peaking at an equivalent
of 2.3 keV and gradually decreasing after that, and an emission area
that remains approximately constant after the rise phase. However,
after 100~s the picture changes: the inferred emission area decreases
sharply by at least an order of magnitude.  The bolometric unabsorbed
fluence in the 300-1500~s time frame is
$3.4\times10^{-8}$~erg~cm$^{-2}$. This is about 3\% of the fluence in
the prompt burst \citep{gal06}.

A similar time-resolved spectral analysis was performed XMM-Newton
data from the 14 low-background bursts, except that this involves an
analysis of 0.7-10 keV photons and channels were binned to make sure
that the number of photons per bin was in excess of 15 to ensure
applicability of the $\chi^2$ statistic (such a procedure was not
necessary for the RXTE data). Prior to the burst spectral modeling we
modeled the pre-burst data to find $N_{\rm H}$ and the photon index to
apply to the burst data. These are $N_{\rm
  H}=3.1\times10^{21}$~cm$^{-2}$ and $\Gamma=1.47$ ($\chi^2_\nu=1.001$
for $\nu=2923$ over two spectra, for a systematic uncertainty per bin
of 2\%). The results of the burst spectra modeling
(Fig.~\ref{figxmmspectralfits}) are generally consistent with the RXTE
results, except at the start and at the end of tail (beyond about 1000
s after burst onset) where a comparison becomes difficult because of
statistical issues. This shows that the drop in radius at 100 s as
seen with RXTE is not related to the lack of low-energy coverage of
that instrument since it is also seen with XMM-Newton for which the
bandpass is extended with the 0.7-4.0 keV photon energy range.

We studied the timing properties of the tail in comparison to those of
the pre-burst data. Fourier power density spectra were generated of
data in the pre-burst -4000/-1000 s time frame and in the tail
+1000/+3000 s time frame. Event mode data were employed at a time
resolution of 2$^{-11}$~s (roughly 0.5 ms), as long as they were
available (see Table~\ref{tab1}). Power spectra were made in 1968 8-s
data stretches for the pre-burst data and averaged, and for 2574 8-s
data stretches in the tail data and averaged. Out of the {\tt FTOOL}
package vs 6.5, {\tt powspec} was employed for this purpose. The
relative difference between these two average power spectra is
presented in Fig.~\ref{figpds}. There is no difference between both
spectra, indicating that nothing significantly changed in the
accretion stream between these periods.

\section{Other bursters}

We checked the average burst profiles of other known prolific bursters
in the PCA data archive, employing the RXTE bursts catalog
\citep{gal06}.  Most of these have variable accretion fluxes and are,
therefore, difficult to analyze for the reasons discussed in the
introduction.  Nevertheless, they sometimes indicate long tails,
although never as long as in \bron.  In this section we report briefly
two cases.

\subsection{4U 1728-34 = GX 354-0}
\label{gx354-0}

4U 1728-34 was observed 346 times for a total PCA exposure time of
1.94 Ms and a total of 106 bursts were detected; a large portion of
these, 69, show photospheric radius expansion. All bursts are short
and have time scales between 4.4 and 8.7~s \citep{gal06}. 40 bursts
have good coverage and are not within too wild accretion flux
variations to search for a tail.  It turns out that the average
profile of these bursts extends to about 800~s which is 10$^2$ times
longer than the initial burst phase. The exponential decay times are
477$\pm$80~s for 2-4 keV, 314$\pm$26~s for 4-9 keV (between 300 and
800 s) and $306\pm$12~s for 2-9 keV.  A power-law fit to the tail in
the latter bandpass between 100 and 800 s yields a decay index of
$1.17\pm0.02$.

We performed time-resolved spectroscopy in a similar manner as for
\bron\ (i.e., the persistent flux as determined from pre-burst data
was subtracted prior to the spectral modelling). There was no need to
include a power law. 36 bursts were selected for this procedure and
only PCU2 data were employed. Since there is no majority within one
RXTE PCA gain epoch, we employed bursts from epochs 3a, 3b, 4 and 5,
averaged them per gain epoch, modeled per gain epoch and averaged the
fitted parameters over the epochs per time bin. A warning is
appropriate here: the bursts in 4U 1728-34 vary more than in \bron;
averaging them will smooth out short features. The results yield the
profiles in Fig.~\ref{fig1728spec} ($\chi^2_\nu$ was averaged as well;
although that number does not adhere to $\chi^2$ statistical
properties, it does give a sense of the overall quality of the fit per
bin). The bolometric flux can be followed downward over 3 decades. It
has a smoother evolution than for \bron, possibly because the
Eddington limit is reached for most bursts so that the flux flattens
against a ceiling representing that limit. Furthermore noticeable is
the drop in radius beyond 10 s, a similar effect as seen in
\bron\ beyond 100 s. The fluence of the tail is 9\% of that of the
prompt emission, taking 100~s as the boundary between prompt burst and
burst tail. We scale the boundary with the decay time $\tau_2$ of the
latter part of the prompt burst \citep[see Table~\ref{tab2}, as read
  from][]{gal06}.

\subsection{EXO 0748-676}

EXO 0748-676 is well known for the accretion-disk edge-on line of
sight, causing eclipses and dips every orbital period of 3.8 hr
\citep{par86}, and also for exhibiting very short burst recurrence
times in certain accretion rate regimes. \cite{boi07} discovered with
XMM-Newton that there are times when this source exhibits 3 bursts in
a row within only half an hour. The first burst in such a `triplet'
always shows a longer tail than the subsequent bursts. In the average
burst profiles the e-folding decay time is about 2.5 times longer for
the first bursts than for the subsequent bursts (50-55 s versus 14-19
s).

RXTE observed EXO~0748-676 ninety-four times for a total exposure of
1.39~Ms. Eighty-four bursts were detected. However, many of these
observations were concentrated on catching eclipses and, thus, involve
only small time stretches rendering burst tail studies
impossible. Also, the bursts are rather weak so that multiple PCUs are
needed to perform meaningful analyses. Rather, we employ XMM-Newton
data to determine average burst profiles.  EXO~0748-676 is the burster
which, with XMM-Newton, was most intensely observed.  In 158 hr of
exposure investigated by \cite{boi07} there are 33 singlet bursts
detected, next to 14 doublets and 5 triplets. We determined average
time profiles of 31 singlets that are not affected by eclipses or dips
within 2000 s, in the same bandpasses as for the RXTE data for
GS~1826-24 and 4U 1728-34.  These also show a long tail. The
exponential decay times are $266\pm122$ s and $260\pm120$~s for 2-4
and 4-9 keV respectively. Again no clear cooling is observed in this
tail.  The equivalent power-law decay index is $2.1\pm0.8$ for 2-4 keV
and $1.6\pm0.8$ for 4-9 keV.

\begin{figure}[t]
\includegraphics[height=\columnwidth,angle=270]{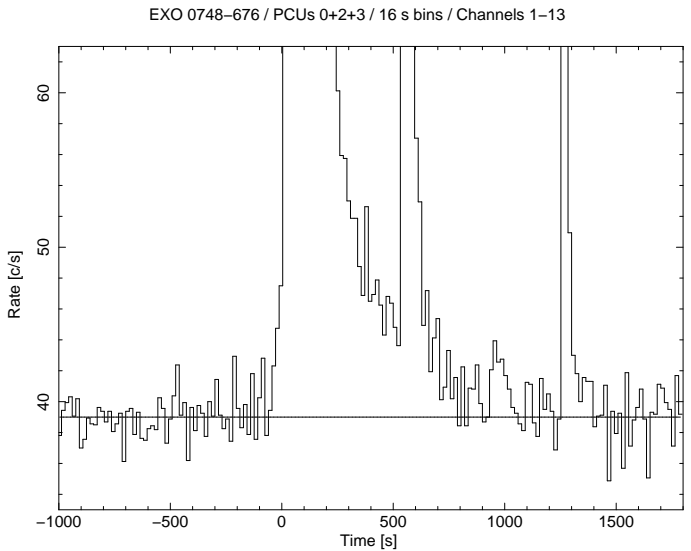}
\caption{Light curve of triple burst from EXO 0748-676 seen on
  September 12, 2006, at 16:30:36 UT (ObsID 92013-01-03-000). The
  peaks of the bursts are truncated to zoom in on the long tail of the
  first burst. The peak fluxes of the 3 bursts are 785.5, 462.1 and
  86.8 c/s. The horizontal line is plotted as a reference for the
  out-of-burst flux.\label{figtriple}}
\end{figure}

\begin{figure}[t]
\includegraphics[width=\columnwidth,bb=64 36 378 515,clip]{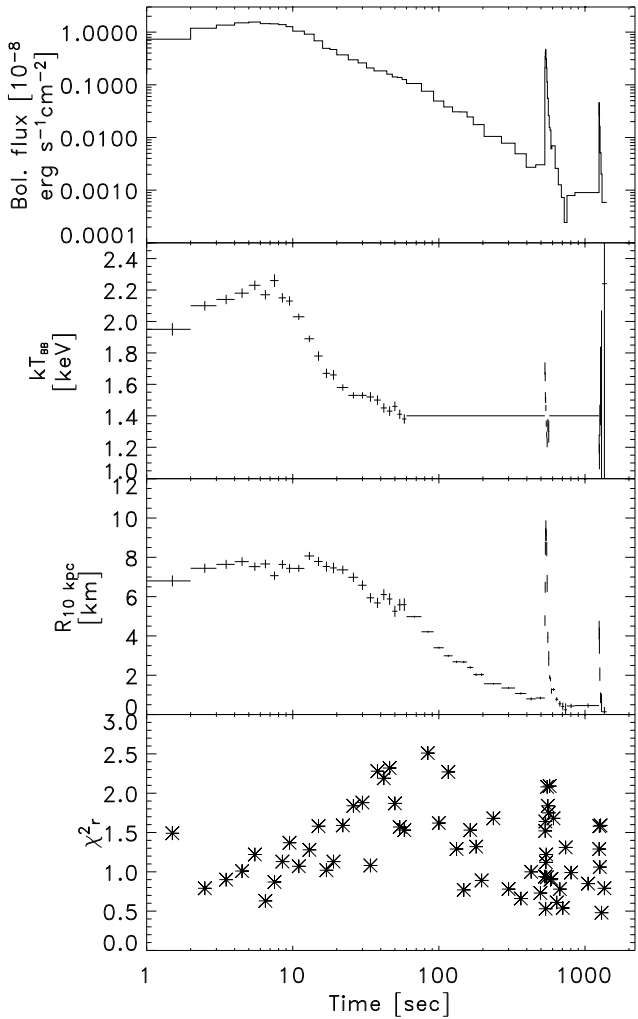}
\caption{Time-resolved spectroscopy of the triple burst from EXO
  0748-676.  A pre-burst spectrum was subtracted before fitting the
  spectra with an absorbed black body model with $N_{\rm
    H}=8\times10^{21}$~cm$^{-2}$. For data between 60 to 1200 s, we
  fixed k$T$ for the tail emission to 1.4 keV, equal to the average
  value if k$T$ is left free for the relevant time intervals.
\label{figexosp}}
\end{figure}

RXTE data do show at least one interesting burst. It is a burst
triplet that occurred on September 12, 2006 (onset of first burst at
16:30:36 UT). The light curve, as extracted from standard 2 data in
ObsID 92013-01-03-000, is presented in Fig.~\ref{figtriple}.
Standard-2 data from PCUs 0, 2 and 3 were added within 2.0-7.3 keV and
the time resolution of these data is 16 s. The results of
time-resolved spectroscopy are shown in Fig.~\ref{figexosp}. The
unabsorbed bolometric fluence ratio over the 3 bursts is 5~:~1~:~1/10
(for the first burst accumulating fluence over only the first 300~s,
so excluding the long tail). The light curve clearly shows a long tail
to the first burst. The e-folding decay time is $300\pm42$~s (measured
between 400 and 1200 s after burst onset and excluding the prompt flux
of the 2nd burst). This is similar as found in the average XMM-Newton
burst profile. The fluence in the tail, excluding the three bursts, is
4.5$\times10^{-9}$~erg~cm$^{-2}$ which is $1.2\pm0.1$\% of the prompt
emission from the first burst (taking 300 s as the boundary between
both).  It is interesting to note that, in the bolometric flux, there
is no clear boundary between the long tail in this triple burst and
the prompt burst, in the sense that there is no excess of the early
emission above the backward extrapolation of the tail. This is
contrast to the situation for \bron\ (cf., Fig. \ref{figxtespectra}
top panel). The same applies to 4U 1728-34 (Fig. \ref{fig1728spec}).

What is most remarkable about this long tail is that it is seemingly
unaffected by the occurrence of the second burst. The tail progresses
undisturbed along the same decay curve. This suggests that the first
burst and its tail emission have a different origin in the NS envelope
than the second burst, either a different layer or a different
locality on the surface. However, this is not testable: the fluence of
the second burst is five times smaller than that of the first. If the
tail would be proportionally smaller, the statistical significance of
it will have dropped below the detection threshold.

\section{Discussion}

Our measurements are summarized in Table~\ref{tab2}. We find that
X-ray bursts from GS 1826-24 show a long tail. In other words, they
exhibit a dual time profile with a prompt burst phase lasting a few
hundred seconds and a tail phase with an e-folding decay time of
10$^3$~s, a flux level of less than $\approx$1\% of the peak flux and
a fluence that is 3\% of that of prompt burst. Beyond the first 200 s
the spectrum can be modeled by a black body with a very slowly
decreasing temperature of $\approx$0.9 keV and a strongly decreasing
emission area.  We find similar tails in RXTE data of 4U 1728-34 and
RXTE and XMM-Newton data of EXO 0748-676, although the contrast
between prompt and tail phase is less pronounced in those
cases. Generally the detection of tails is difficult because of
confusion with variable accretion radiation. Still, bright pronounced
tails have been reported in the literature for a few individual
bursts. The two most obvious questions about the long tails are: what
is the physical cause and why is hardly any cooling detected? We
investigate the first question in Sect.~\ref{dis1}, keeping in mind
only \bron. In Sect. \ref{dis2} we touch on the second question,
considering 4U 1728-34 and EXO 0748-676 as well. In Sect. \ref{other}
we compare \bron\ with the other cases listed in Table~\ref{tab2}.

\begin{table*}
\begin{center}
\caption[]{Summary of measurements\label{tab2}}
\begin{tabular}{|l|l|l|l|l|l|l|}
\hline\hline
Burst source & \multicolumn{3}{|c|}{Mean burst duration (s)$^a$} & \multicolumn{2}{|c|}{Tail/prompt ratio}  & Ref.$^b$ \\
        & time scale$^c$ & $\tau_2^d$ & tail$^e$ & in time$^f$ & in fluence$^g$ & \\
\hline
\multicolumn{7}{|c|}{\it Sources discussed here} \\
\hline
GS 1826-24    &$39.4\pm0.2$     & 44.6 & $1252\pm25$         & 32/28 & 0.03 (300 s) & \\
              &                 &      & {\it (600-5500 s)}  &       &              & \\
              &                 &      &$\chi^2/\nu$=16.9/15 &       &              & \\
4U 1728-34    & $5.85\pm0.02$   & 12.6 &  $313\pm17$         & 53/25 & 0.09 (100 s)  & \\
              &                 &      &  {\it (200-800 s)}  &       &              & \\
              &                 &      &$\chi^2/\nu$=13.4/11 &       &              & \\
EXO 0748-676  & $13.7\pm0.1$    & 36.3 & $300\pm42^h$        & 22/8  & 0.01 (300 s) & \\
              &                 &      & {\it (400-1200 s)}  &       &              & \\
              &                 &      &$\chi^2/\nu$=141/44  &       &              & \\
\hline
\multicolumn{7}{|c|}{\it Literature cases} \\
\hline
GX 3+1        & $3\pm1$         &  6   & 1110$\pm170$&370/185& 40 (10 s) & 1 \\
M28           &                 & 7.5  & 800-3250    & $\sim100$& $>1.4^i$ (90 s)& 2\\
Aql X-1       & $18.4$          &      & $627\pm100$ & 34    &    & 3 \\
OSO-8 long burst& 44.8          &      & 215.7       &  5    & 0.01 (150 s) & 4 \\
\hline\hline
\end{tabular}\\
$^a$Averaged over the bursts cataloged by \cite{gal06}; $^b$References: 1 - \cite{che06}, 2 - \cite{got97}, 3 - \cite{cze87}, 4 - \cite{swa77}; $^c$The time scale is defined as the burst fluence divided by the peak flux; $^d$$\tau_2$ is the e-folding decay time of the final part of the prompt burst; $^e$The time range in parentheses refers to the interval in which the fit was carried out; $^f$The two numbers refer to division by the mean time scale and $\tau_2$, respectively; $^g$ Fluences for GS 1826-24 and 4U 1728-34 were determined from data averaged over multiple bursts; that for EXO 0748-676 from the RXTE triple burst. The time between parentheses refers to the interval for the prompt emission; $^h$this value applies to the triple burst only and excluding times for the second burst ; $^i$based on counts of 0.8-12 keV photons in ASCA-GIS \citep[table 1 in][]{got97}
\end{center}
\end{table*}

\subsection{Origin of the long tail}
\label{dis1}

\begin{figure}[t]
\includegraphics[width=\columnwidth,bb=20 0 468 484,clip,angle=0]{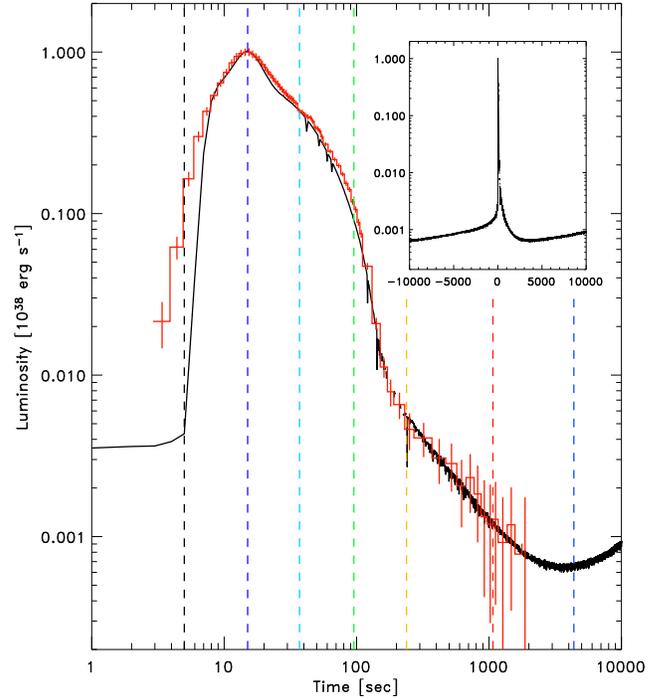}
\caption{Average observed (red histogram) and predicted (solid) time
  profiles of bolometric flux, after normalization at the respective
  peak values and alignment at the peak. The vertical dashed lines
  indicate the times for the depth profiles plotted in
  Fig.~\ref{figcycles}. The inset shows the model on a broader and
  linear time scale, exhibiting the slow luminosity increase between
  bursts from burning of increasing amounts of accreted hydrogen in
  the hot CNO cycle.\label{figmodel}}
\end{figure}

What is the origin of the 10$^3$ s time scale of the long tail? One
idea spawns from the detection of a hot plasma surrounding the NS
\citep{tho05}, namely that prompt burst photons are trapped in the hot
plasma through scattering and that the 10$^3$~s time scale is the time
it takes to drain the plasma of those photons. The maximum size of the
plasma in a 2.1 hr orbit \citep{hom98} is of order 10$^{11}$~cm. For
an optical depth of 6 \citep{tho05}, this implies a drainage time of
at most 10-10$^2$ s. This is one to two orders of magnitude shorter
than observed. Therefore, if 2.1 hr is indeed the orbital period
\citep[this needs to be verified; e.g.,][]{mes04}, the 10$^3$~s time
scale cannot be explained by this idea.

Another idea is that the tail is due to cooling of layers that are
deeper than the flash layer.  In a thermonuclear shell flash, heat is
transported upward (to be radiated by the photosphere) as well as
downward \citep[heating up deeper layers through conduction;
][]{eich89,stro02}.  Deeper layers, extending down to the crust,
consist of the ashes of the nuclear H and He burning and are rich in
elements up to a mass number between 60 and 100
\citep{sch01,fis08}. Radiative cooling of these deeper layers can
explain the long tail.  The time scale of the long tail points to a
column depth that is 10 to 30 times larger than that of the burning
layer. The amount of fluence in the tail is one to two orders of
magnitude smaller than in the prompt burst, suggesting less heating
downward than upward.

\begin{figure*}[t]
\includegraphics[angle=0]{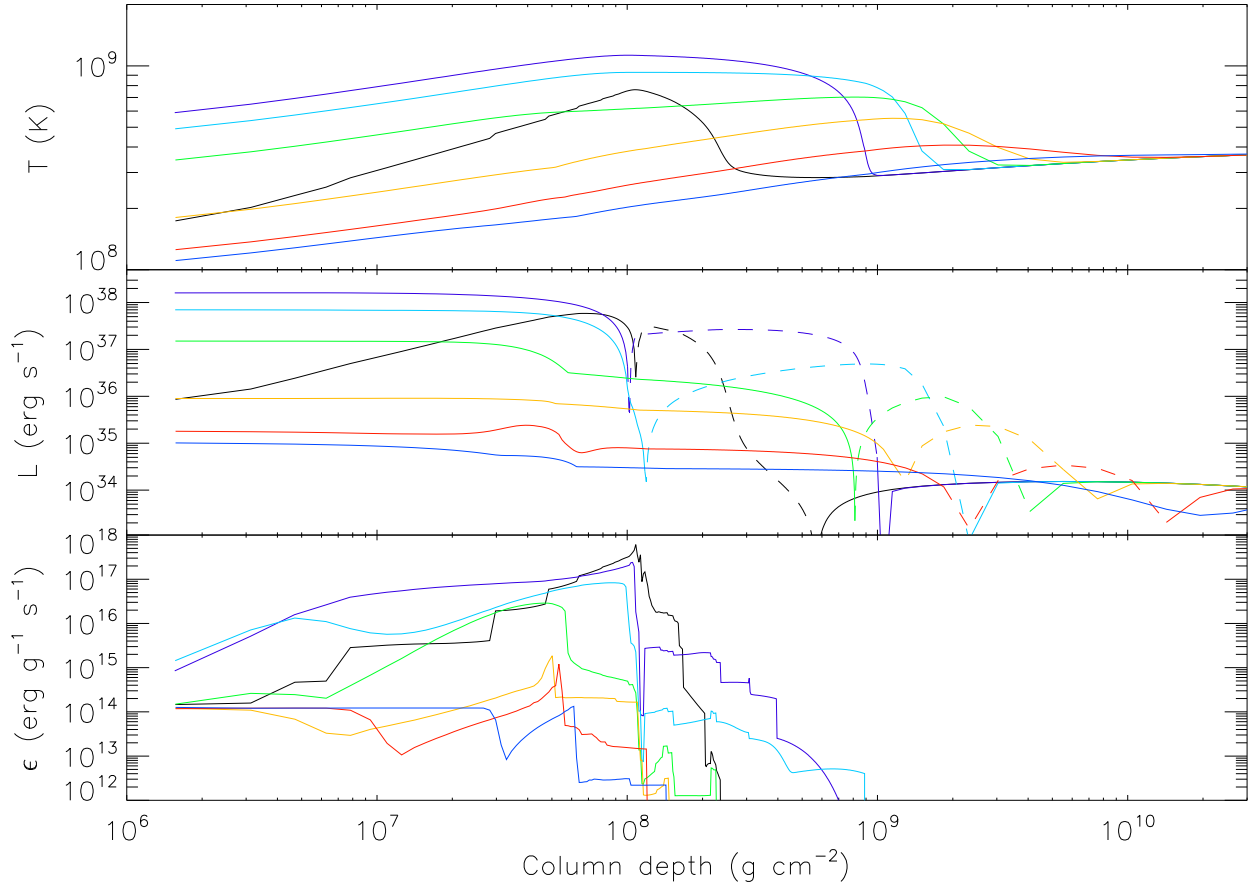}
\caption{Profiles of temperature (top panel), net outward luminosity
  (middle; dashed parts indicate inward luminosity) and specific
  nuclear energy generation rate (bottom) calculated as a function of
  depth for a flash model specifically for \bron\ \citep[model `A3'
    of][]{heg07}. Seven profiles are shown. The black curve is for
  $t=0$ (burst onset), the other curves are for the times indicated
  with vertical lines in Fig.~\ref{figmodel} with the same color. The
  top panel shows the inward heating (compare with the location of the
  heat source in the bottom panel, down to only
  $9\times10^8$~g~cm$^{-2}$) that is thought to be responsible for the
  long tails on the X-ray bursts of \bron. See also
  Fig.~\ref{figheger}.\label{figcycles}}
\end{figure*}

\begin{figure}[t]
\includegraphics[width=\columnwidth,angle=0]{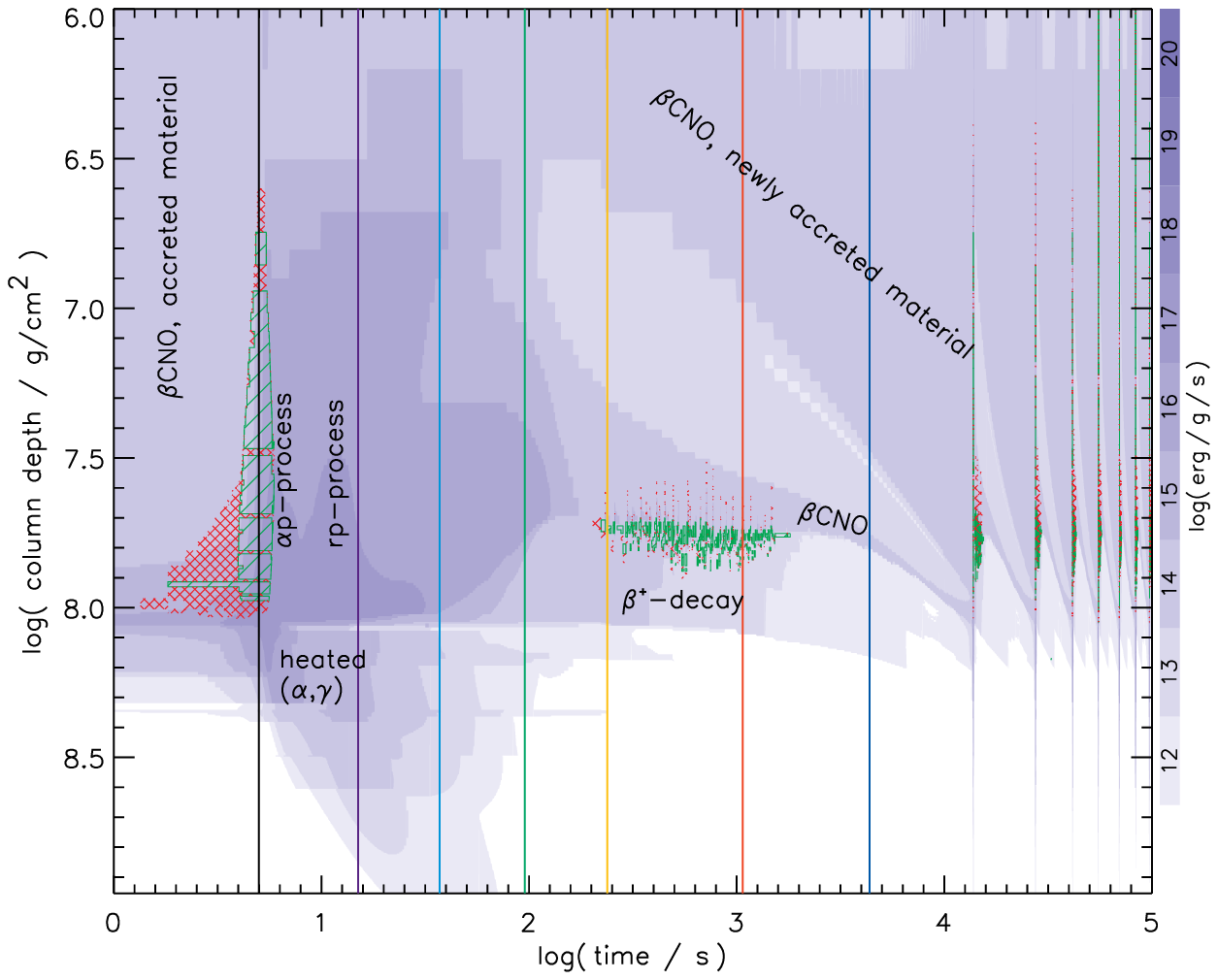}
\caption{Dynamic depth profile of the specific nuclear energy
  generation rate, as calculated in model `A3' \citep{heg07}. Each
  level of shading indicates a change in the rate by one order of
  magnitude. The bottom of this plot is taken to be the bottom of the
  reservoir of ashes in the model. The hatched regions indicate
  convective layers \citep[green hatched for convection and red
    cross-hatched for semiconvection; see][]{woo04}. The convective
  region between 10$^2$ and 10$^3$ s is located at the bottom of the
  hydrogen left over after the burst and is probably due to a
  Rayleigh-Taylor instability related to composition inversion
  resulting from the burning. It does not have a major influence on
  the structure and evolution of the model. `$\beta$CNO' refers to a
  variety of the CNO cycle where the reaction rates are limited by
  $\beta$ decays; this is the `hot' CNO cycle. `$(\alpha,\gamma)$'
  refers to $\alpha$ capture by heavier isotopes such as $^{12}$C and
  $^{16}$O. A series of bursts is shown; the next burst occurs at
  log(time/s)=4.15.\label{figheger}}
\end{figure}

The idea that the long tail results from cooling of deeper layers is
corroborated by model calculations.  \cite{heg07} calculated various
sequences of flashes specifically for \bron\ with different mass
accretion rates and metallicities. Their model `A3' is the one whose
recurrence time of 3.85~hr matches best the majority of our
bursts. This model assumes a mass accretion rate of
$1.58\times10^{-9}$~$M_\odot$~yr$^{-1}$ (or
$1\times10^{17}$~g~s$^{-1}$) and a metallicity of $Z=0.02$. The
long-term time profile of the radiated luminosity is very similar to
Fig.~27 of \cite{woo04}. Ignoring the period before the first burst
and after the accretion turn-on, the interburst time profile is
characterized by a gradual decline till about 3000 s after the burst
onset followed by a gradual increase for about 10$^4$ s untill the
next burst.  The decline is due to cooling of the deeper layers; the
increase is due to hot CNO burning of newly accreted hydrogen in the
burning zone. In Fig.~\ref{figmodel} we plot the observed bolometric
flux (see also Fig.~\ref{figxtespectra}) and the average luminosity
profile of 29 bursts from the `A3' model by \cite{heg07}. The fluxes
and luminosities were normalized to the peak value, and the light
curves were aligned at the peak. Times in the model were corrected for
general relativistic effects \citep[through a multiplication with
  $1+z=1.26$;][]{woo04}. The time profiles are an excellent match, all
the way to the long tail. Figure~\ref{figcycles} shows for this model
the evolution of the depth profiles of temperature, net outward
luminosity and nuclear energy generation rate per unit
mass. Figure~\ref{figheger} presents a more detailed view of the
latter panel, showing the dynamic depth profile of the specific
nuclear energy generation rate, annotated with the various nuclear
processes playing a dominant role at the various locations. Most of
the nuclear burning occurs for column depths $y <
2\times10^8$~g~cm$^{-2}$, but layers are heated that are 10 times
deeper (see first panel of Fig.~\ref{figcycles}). The energetically
less important nuclear burning below $2\times10^8$~g~cm$^{-2}$
(`heated $(\alpha,\gamma)$' or heat-induced $\alpha$ capture by
heavier isotopes) is actually the result of conductive heating from
the shallower layers.

The inward heating is due to conduction. To first order, the heat
transport scales with $A/Z^2$ \citep{yu80,bil95,cum01} where $Z$ is
nuclear charge and $A$ nuclear mass number. Therefore, one may expect
less inward heating for a layer with heavier isotopes.  The
composition depends, on its turn, on the composition of the donor
atmosphere and the accretion rate. Thus, one may expect more inward
heating in for instance ultra-compact X-ray binaries or for accretion
rates and burst regimes with high $\alpha$ in which most hydrogen is
burnt through the hot CNO cycle instead of the rp process, because the
rp process produces the heaviest elements. Unfortunately, the $A/Z^2$
proportionality of the heat transport is only a crude approximation so
that inferences are not straightforward, for instance by comparing
different bursters or bursts from the same source at different
accretion rates.

There are other dependencies of the conductivity as well, such as on
ignition depth which may vary from source to source and burst to
burst. This will not only have an effect on the duration of the tail,
but also on the fluence ratio between the tail and the prompt burst.
For a relatively shallow ignition, the inward heating will not go as
deep and the tail will be short and less fluent. This may explain
qualitatively why only the first burst in the triplet from EXO
0748-676 has a long tail that is unaffected by subsequent bursts. The
subsequent bursts ignite at shallower depths.

\begin{figure}[!t]
\includegraphics[height=\columnwidth,angle=270]{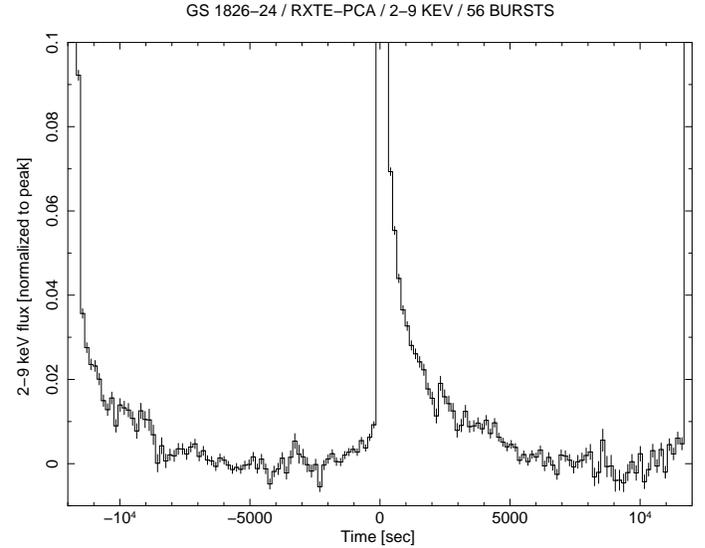}
\caption{Average time profile of 56 X-ray bursts from GS 1826-24 as
  measured with PCU2 on RXTE/PCA in 2-9 keV, now zooming out
  to times further away from the burst.\label{fig1826rxtelong}}
\end{figure}

As mentioned above, the `A3' model predicts a trend in the NS
luminosity between bursts consisting of a gradual decline for 3000 s
after a burst followed by a gradual increase for 10$^4$~s up to the
next burst (see Fig.~\ref{figmodel}). In practice it is difficult to
disentangle the NS luminosity from the flux measurements, but it
interesting to show the straightforward observed photon rate, see
Fig.~\ref{fig1826rxtelong}. This is the same kind of measurement as
shown in Fig.~\ref{figxtefit}, except that it is shown linearly and
between 12,000 s before and after the burst time.  To obtain as much
data as possible, particularly far away from the bursts when the data
coverage is less than close to the bursts, we included 27 additional
bursts for a total of 56. This plot does not show a clear increasing
trend except for the final 2$\times10^3$~s before the next burst. A
similar behavior, but of worse statistical quality, is apparent from
XMM-Newton data. One would expect a linearly increasing trend for the
hot CNO cycle according to $L=4\times10^{34} (Z_{\rm CNO}/0.02)
\dot{M}_{17} t_{\rm hr}$~\lum, with $Z_{\rm CNO}$ the CNO abundance,
$\dot{M}_{17}$ the mass accretion rate in 10$^{17}$~g~s$^{-1}$ and
$t_{\rm hr}$ time in hours, because the amount of accumulated fuel
grows linearly with time and the hot CNO burning rate is a constant
that depends only on the CNO mass fraction \citep{hoy65}. The maximum
slope consistent with the data shown in Fig.~\ref{fig1826rxtelong} is,
for $\dot{M}_{17}=1$ \citep{heg07} and assuming a black body
temperature of 0.5 keV, equivalent to an upper limit of $Z_{\rm
  CNO}<0.05$.

A third idea for explaining the long tails is that the X-ray bursts
influence the accretion disk in such a way that the accretion rate is
temporarily increased. A change in accretion rate by X-ray bursts has
been seen before. For instance, X-ray bursts from 4U 1820-303
\citep{stro02} and 4U 1724-307 \citep{mol00} are so luminous that
radiation pressure blows away the inner parts of the accretion disk,
shutting off accretion for a few seconds. Secondly, there are
suggestions in X-ray bursts from Cen X-4, XTE J1747-214 and 2S
1711-337 that X-ray bursts act as triggers for switching accretion
disks from a cold neutral state to a hot ionized state a few days
later \citep[although that cannot be explained yet in a quantitative
  manner;][]{kuu08}. Lastly, it appears that immediately before and
after superbursts the persistent flux from the accretion disk behaves
differently in the sense that the flux has a somewhat decreased level
for half a day before and an increased level for approximately a day
after superbursts \citep[e.g.,][]{cor00,cor02,kuu02,kee08}. Perhaps
there is a 4th type of effect from X-ray bursts on disks resulting in
long tails. However, `our' long-tailed X-ray bursts are less energetic
than the aforementioned bursts that are either super-Eddington (i.e.,
relatively high flux) or superbursts (high fluence). The X-ray bursts
from \bron\ are not particularly luminous (none show photospheric
radius expansion) nor extraordinarily fluent \citep{gal07}. The same
applies to EXO 0748-676 \citep[only one burst shows photospheric
  radius expansion;][]{wol05}. The majority of the bursts from 4U
1728-34 do show radius expansion, but have very small fluence due to a
very short duration \citep{gal03}. Finally, the power density spectra
before the burst and during the tail are not significantly different
(c.f., Fig.~\ref{figpds}), suggesting no change in accretion stream or
rate. We believe that an explanation for the long tails in terms of a
changed accretion environment is less likely.

\subsection{Lack of strong cooling, decreasing black body area}
\label{dis2}

The lack of cooling could be explained by Compton up-scattering of all
photons by the hot plasma that surrounds the NS and the temperature
inferred from the spectrum may in fact be representative of the plasma
rather than the NS. A problem of this explanation is that the inferred
emission area drops so suddenly in both \bron\ and EXO 0748-676, after
100 and 60 s respectively. The temperature of the burst at that time
is still sufficiently high that one should detect large numbers of
unscattered photons. A decline of the normalization, if at all, is
expected to be more gradual.

Another, in our opinion more likely, explanation is that the neutron
star may already be fairly hot without the heating by thermonuclear
flashes. The effective temperature can then never drop below the
`quiescent' NS value. There is sufficient persistent energy production
to sustain a NS hot enough to explain our measurements, by stable
hydrogen burning via the hot CNO cycle (the model depicted in
Fig.~\ref{figcycles} predicts about $1\times$10$^{35}$~\lum),
pycnonuclear reactions and electron capture processes in the crust
(the same model assumes $1.6\times10^{34}$~\lum\ or 0.15 MeV/nucleon).
Gravitational energy release by the settling of the accreted processed
matter in the envelope is negligible for these accretion rates
\citep{bb98}. Gravitational energy release by accretion occurs just
outside the NS, is radiated away from the NS \citep{kin95}, and is
decoupled from the burst emission as long as the burst flux is less
than the Eddington limit.  The total energy production rate depends on
the mass accretion rate and the H and CNO abundance of the donor
atmosphere. Also, it may partly heat up the core instead of the
photosphere. Luminosities are expected to reach up to at least a few
times 10$^{35}$~\lum. For a canonical 10 km NS radius and ignoring the
gravitational redshift of a few tens of percents, the Stefan-Boltmann
law predicts for a luminosity of 2$\times10^{35}$~\lum\ an effective
temperature of the non-bursting NS of order 0.4 keV. If a burst
occurs, the photosphere temperature rises to a few keV and the
emission is completely dominated by this extra heating, but if the
temperature becomes comparable to the quiescent value, in the tail of
the burst, the spectrum is strongly affected by the already hot NS. In
a standard time-resolved spectroscopic analysis, where a pre-burst
spectrum is subtracted and the net spectrum is modeled through a black
body, one actually subtracts one Planck function from another. The
resulting function is {\em not} a Planck function and if,
nevertheless, it is modeled as such, the physical meaning of inferred
emission areas is lost. This effect has been extensively studied by
\cite{jvp86}. They find 1) that it is particularly important in the
tail of an X-ray burst; 2) that the fit results in constant
temperatures and decreasing emission areas, as we find in our
analyses, and 3) that the derived temperature has a tight relationship
with the NS temperature: the measured temperature is about 30\% higher
than the NS temperature outside bursts. Taken at face value, our
measurements imply a NS temperature of 0.7-0.8 keV for both \bron\ and
EXO 0748-676, ignoring again gravitational redshift and deviations
from black body radiation that become more significant towards lower
temperatures \citep[e.g.,][]{zav96}. This is encouragingly close to
the simple prediction done above of 0.4 keV. A similar effect in
reversed time order may be happening during the rise phase of bursts
from \bron.

\cite{kuu02b} investigate this effect in detail for bursts from the
high-$\dot{M}$ system GX 17+2. \cite{jvp86} suggest to study burst
data without subtracting the pre-burst spectrum and employing a model
that includes components for the accretion disk emission and a black
body for the thermal emission from the NS. \cite{kuu02b} follow this
suggestion and model the accretion disk emission by a cutoff power
law. They find unacceptable values for the goodness of fit and dismiss
the explanation by an already hot NS and put forward the possibility
that the decreasing black body radius is connected to blanketing
effects in the NS atmosphere and comptonization of burst photons in
the NS atmosphere. Theoretical calculations
\citep{lon86,ebi87,pav91,zav96,maj05} show that the color temperature
is between 1.2 and 1.7 times the effective temperature.  The fitted
black body radius thus decreases by the square of that, to maintain
the same bolometric flux. This cannot explain the drop in radius that
we observe which is at least a factor of 5.

As \cite{kuu02} point out, the reason for the unsuccesful modeling for
GX 17+2 is that the accretion disk spectrum also contains a strong
black body component due to the high $\dot{M}$ in this LMXB. The disk
black body has an only slightly higher temperature than that expected
of the NS and therefore the latter is difficult to distinguish. Also,
in GX 17+2 the low-energy absorption is high with $N_{\rm
  H}=1.9\times10^{22}$~cm$^{-2}$ \citep{far07} so that it is tough to
find evidence for a NS of temperature k$T\approx0.5$ keV. Finally, the
flux is expected to be low, of order 0.1\% of the burst peak, so that
accumulating a statistically relevant spectrum is challenging. Our
measurements of \bron\ do not suffer from these difficulties. Perhaps
a decreasing radius and a flattening temperature in a `standard' burst
analysis (i.e., with subtraction of the pre-burst spectrum) constitute
the best possible evidence for a hot NS.

We note that many X-ray bursts in the RXTE catalog \citep[][]{gal06}
show similar behavior: temperatures remaining above $\approx$1 keV and
decreasing fitted radii. Since an explanation by a NS that is already
hot without flashes is more likely to be applicable to a major portion
of the burster population than scattering in a hot circumstellar
plasma, this provides additional support to that explanation.
However, these data are vulnerable to lack of low-energy coverage of
the PCA, as well as from sometimes high absorption columns. As a
result, it is quite difficult to accurately measure k$T$ below 1
keV. \bron\ is one of the few cases where this problem does not exist:
we have the XMM-Newton data to corroborate the RXTE data and a low
$N_{\rm H}$.

We furthermore note that the behavior of burst tails in UCXBs is
expected to be markedly different in the hot-NS scenario for two
reasons. Firstly, UCXBs have much lower H abundances so that hot CNO
burning provides much less heating outside flashes. Secondly, in many
UCXBs the accretion rate is lower so that again the energy production
rate is lower outside bursts and the NS cooler. This expectation is in
line with high-sensitivity spectroscopy of some X-ray bursts from
UCXBs, for example in A~1246-588 \citep[][]{zan08} where the
temperature is seen to decay to 0.5 keV.

\subsection{Long tails in other sources}
\label{other}

\begin{figure}[t]
\includegraphics[height=\columnwidth,angle=270]{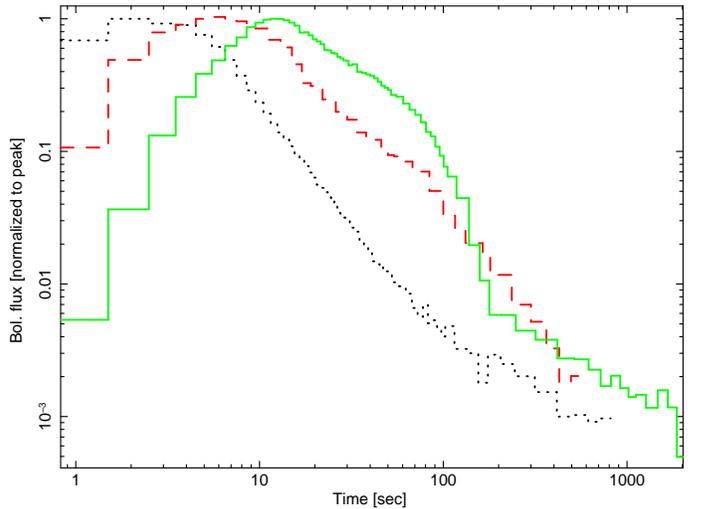}
\caption{The average time profiles of the bolometric flux for
  \bron\ (solid green curve) and 4U 1728-34 (dotted black) and the
  profile for the triple burst from EXO 0748-676 (dashes red; the
  profile is cut at the time the 2nd burst occurs). Fluxes are
  renormalized with the respective peak values.\label{figbolflux}}
\end{figure}

Figure~\ref{figbolflux} shows the time profiles of the bolometric flux
for 4U 1728-34 and EXO 0748-676, together with that for \bron. These
are the same data as shown in the top panels of
Figs.~\ref{figxtespectra}, \ref{fig1728spec} and \ref{figexosp},
except that one earlier time bin is shown. For convenience the fluxes
have been normalized to the respective peak values. This figure shows
that there is one property that distinguishes \bron\ from the other
two sources: there is a clear difference between the prompt and the
tail phase. The tails of 4U 1728-34 and EXO 0748-676 are smooth
extensions of the initial burst phase. This may be related to the
energetic contribution of the rp-process being larger in \bron\ (cf,
Fig.~\ref{figheger}). It is probably not related to a smaller amount
of inward heating in 4U 1728-34 and EXO 0748-676. The alternative
explanation, deeper ignition depths in 4U 1728-34 and EXO 0748-676, is
not consistent with the small burst recurrence times, similar to
\bron\ \citep{gal06}.

The other documented cases \citep[][see
  Table~\ref{tab2}]{swa77,cze87,got97,che06} do seem to have dual time
profiles. Their higher tail-to-peak flux ratio indicates that the
heating to the deeper layers can be even more efficient than seen in
\bron. Perhaps there is a smaller abundance of heavy
isotopes. However, this is difficult to assess quantitatively because
a quantitative comparison with theoretical models has yet to be
carried out.

An exceptional case is the long tail in GX 3+1 \citep{che06}. It
contains much more fluence than the prompt burst, perhaps 40
times as much\footnote{this value depends on where the boundary
  between prompt burst and tail is chosen, but it is always
  significantly larger than 1}! Aql X-1 \citep{cze87} and the source
in M28 \citep{got97} may also have large fluence ratios but there are
data gaps that preclude verification. If interpreted as direct NS
emission, this cannot be explained through cooling of deep layers,
particularly since the prompt fluence is similar to that for other
bursts from GX 3+1 that do not show a long tail. One difference that
distinguishes these other cases from the cases discussed in the paper
is that they were most probably in a different burst regime when the
long-tailed bursts occurred, namely in a regime without continuous
hot-CNO hydrogen burning.

Several classes of long X-ray bursts, with e-folding decay times in
excess of roughly 100 s, have been discovered in the past decade. This
includes superbursts \citep[flashes of 100 m thick carbon-rich layers;
  for a recent list, see][]{kee09} and intermediately long bursts that
may result from flashes of 10~m thick helium layers
\citep{zan05a,cum06,zan07} or otherwise \citep{che07,lin08}. With the
present paper, one realizes that even the `classical' short X-ray
bursts can be similarly long in some sense.  The distinguishing factor
is that these bursts initiate with a normal-sized X-ray burst and
that, probably, the implied thick layer is not heated locally by
nuclear reactions but by conduction from a hotter layer on top.

\section{Conclusion}

We have detected an hour-long tail to bursts in \bron\ with fluxes and
fluences that are two orders of magnitude smaller than those of the
bursts themselves. We have found similar tails in bursts from 4U
1728-34 and EXO 0748-676, although they are less distinguished from
the prompt burst emission. While detection in other bursters is
hampered by varying accretion fluxes of similar magnitude, there are
reports of individual cases of bursts with long tails, most notably in
GX 3+1 \citep{che06}. Model calculations show that the tail in
\bron\ can be explained by delayed cooling of layers that are up to
ten times deeper in column density than where the flash occurs and
that were heated up through inward conduction of the flash
heat. Possibly tails in other sources can be similarly explained.
Further model calculations are needed, where dependencies of donor
composition, ignition depth and accretion rate are taken into account.
Comparing such calculations to different kinds of bursts and bursters
may yield constraints on the details of conduction in the neutron star
envelope.

A characteristic of the tails, that at first hand is unexpected in
this scenario, is the small amount of cooling. Rather than being due
to up-scattering of the burst photons by a hotter optically-thick
plasma, we believe that the most likely explanation is that the NS is
already hot without flashes. The temperature that one measures in the
burst tail is then representative for the non-bursting NS. This
scenario also provides a more natural explanation for the decreasing
black body normalization in the tails. As discussed by \cite{jvp86},
this presents an interesting opportunity to study NS temperatures as a
function of accretion rate. However, this may be a cumbersome. Apart
from good low-energy coverage, one would need to work on a
single-burst basis, also for the accurate modeling of the persistent
spectrum. Better data would be needed than presented here.

\acknowledgements We are grateful to Duncan Galloway and Peter Jonker
for useful discussions and an anonymous referee for a swift review and
useful suggestions. LK was supported by the Netherlands Organization
for Research (NWO) through NOVA, AC by an NSERC Disovery Grant, Le
Fonds Qu\'{e}b\'{e}cois de la Recherche sur la Nature et les
Technologies and the Canadian Institute for Adcanced Research. AC is
an Alfred P. Sloan Research Fellow. AH was supported by the DOE
Program for Scientific Discovery through Advanced Computing (SciDAC;
DOE-FC02-01ER41176 and DOE-FC02-06ER41438) and by the US Department of
Energy under grant DE-FG02-87ER40328.

\bibliographystyle{aa} \bibliography{references}

\end{document}